\documentstyle[12pt]{article}

\global\arraycolsep=2pt
\input{epsf}

\begin{document}

\thispagestyle{empty}
\begin{titlepage}

\begin{flushright}
CERN-TH.7156/94\\
hep-ph/9402288
\end{flushright}

\vspace{0.3cm}

\begin{center}
\Large\bf Resummation of Nonperturbative Corrections to the Lepton
Spectrum in Inclusive $B\to X\,\ell\,\bar\nu$ Decays
\end{center}

\vspace{0.8cm}

\begin{center}
Thomas Mannel and Matthias Neubert\\
{\sl Theory Division, CERN, CH-1211 Geneva 23, Switzerland}
\end{center}

\vspace{0.8cm}

\begin{abstract}
\noindent
We apply the operator product expansion to resum the leading
nonperturbative corrections to the endpoint region of the lepton
spectrum in inclusive semileptonic $B\to X_q\,\ell\,\bar\nu$ decays,
taking into account a finite quark mass $m_q$ in the final state. We
show that both for $b\to c$ and $b\to u$ transitions, it is
consistent to describe these effects by a convolution of the parton
model spectrum with a fundamental light-cone structure function. The
moments of this function are proportional to forward matrix elements
of higher-dimension operators. The prospects for an extraction of the
structure function from a measurement of the lepton spectrum are
discussed.
\end{abstract}

\centerline{(submitted to Physical Review D)}
\bigskip\bigskip

\noindent
CERN-TH.7156/94\\
February 1994

\end{titlepage}

\section{Introduction}

The heavy quark expansion, i.e.\ a systematic expansion in powers of
$\Lambda/m_Q$ (we use $\Lambda$ as a generic notation for a typical
hadronic scale of the strong interactions), has by now become a
widely used tool in the theoretical description of systems containing
a heavy quark $Q$ interacting with light degrees of freedom
\cite{Eich,Geor,Mann,Falk}. Its application to exclusive decays of
heavy mesons and baryons has been explored already in great detail
\cite{habil}. Recently, the idea has been put forward to generalize
the heavy quark expansion to obtain a QCD-based description of
inclusive decays of heavy hadrons
\cite{Chay,Bigi,Blok,MaWe,Adam,Thom}. Similarly to the case of
deep-inelastic scattering, an operator product expansion is applied
to the product of two local currents. The mass of the decaying quark
provides the large momentum scale. The total decay rates can then be
written as an expansion in inverse powers of $m_Q$. The operators
appearing at leading order have dimension three and correspond to the
free-quark decay. The matrix elements of the dimension-four operators
vanish by the equation of motion, and thus the leading
nonperturbative corrections arise from dimension-five operators and
are of order $(\Lambda/m_Q)^2$. For the case of $B$-meson decays,
they can be parameterized by two low-energy parameters $\lambda_1$
and $\lambda_2$, which are related to the kinetic energy $K_b$ of the
$b$-quark inside the $B$-meson, and to the mass splitting between
$B$- and $B^*$-mesons. They are defined as \cite{FaNe}
\begin{equation}\label{lamdef}
   K_b = {\langle\,\vec p_b^{\ 2}\rangle\over 2 m_b}
   = - {\lambda_1\over 2 m_b} \,,\qquad
   m_{B^*}^2 - m_B^2 = 4\lambda_2 \,.
\end{equation}

The operator product expansion has also been used to calculate
differential distributions, such as the charged-lepton energy
spectrum in inclusive semileptonic decays. In this case, the relevant
large momentum scale is $Q=m_b v-q$, where $q$ denotes the momentum
transfer to the lepton pair, and $v$ is the velocity of the decaying
hadron. After integrating over the neutrino momentum, the operator
product expansion is a combined expansion in powers of $\Lambda/m_b$
and $\Lambda/(m_b-2 E_\ell)$, where $E_\ell$ is the lepton energy in
the rest frame of the decaying hadron. Over most of the available
phase space these parameters are of similar magnitude. However, close
to the endpoint, i.e.\ for $(m_b-2 E_\ell)$ of order $\Lambda$, the
second expansion parameter is of order unity, and a partial
resummation of the expansion becomes necessary. For the lepton
spectrum in charmless semileptonic $B\to X_u\,\ell\,\bar\nu$ decays,
as well as for the photon spectrum in the penguin-induced $B\to
X_s\,\gamma$ decays, this resummation has been constructed in
Refs.~\cite{shape,photon,Fermi}, neglecting the mass of the quark in
the final state. It has been shown that the leading nonperturbative
effects can be related to a universal structure function, which
describes the distribution of the light-cone residual momentum of the
heavy quark inside the decaying hadron. The purpose of this paper is
to generalize this approach to the case where the mass of the quark
in the final state cannot be neglected. For simplicity, we will work
to leading order in perturbation theory. Radiative corrections, which
have been calculated at the parton level in
Ref.~\cite{AlPi,CCM,JeKu}, should however be included before
confronting our results with experimental data. We shall consider
$B\to X_q\,\ell\,\bar\nu$ transitions and treat $m_q$ as a free
parameter. As we shall see, the presence of several mass scales leads
to technical and conceptual complications. Let us define the
dimensionless ratio
\begin{equation}\label{rhodef}
   \rho = {m_q^2\over m_b^2} \,.
\end{equation}
It will be natural to distinguish the three cases where $\rho$ is of
order 1, $\Lambda/m_b$, and $(\Lambda/m_b)^2$, respectively. The
first case is of no phenomenological importance and will not be
considered in detail. The second case is relevant for $b\to c$
transitions, where $\rho\simeq 0.1$. The third case applies, e.g.,
when one studies the effect of a small constituent mass of the
$u$-quark in $b\to u$ transitions.

\begin{figure}[h]
   \vspace{0.5cm}
   \epsfysize=7cm
   \centerline{\epsffile{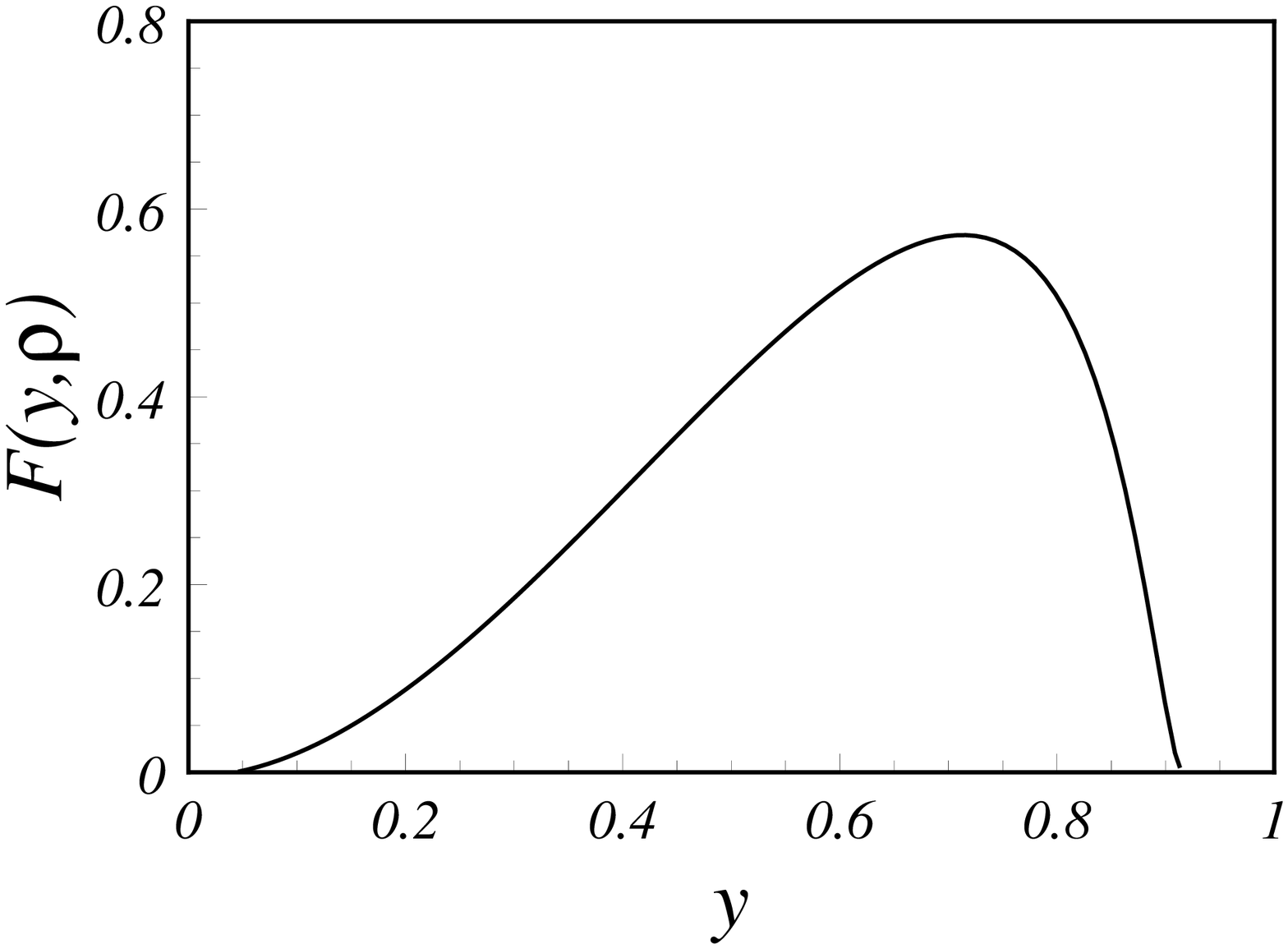}}
   \centerline{\parbox{11cm}{\caption{\label{fig:1}
Shape of the lepton spectrum predicted in the free-quark decay model.
We use $\rho=(m_c/m_b)^2 =0.08$.
   }}}
\end{figure}

The necessity of a resummation of the naive expansion in powers of
$\Lambda/m_b$ close to the endpoint is apparent from the result
obtained in Refs.~\cite{Bigi,Blok,MaWe} for the lepton spectrum in
$B\to X_q\,\ell\,\bar\nu$ decays. Let us define
\begin{equation}
   y = {2 E_\ell\over m_b} \,,\qquad
   \Gamma_b = {G_F^2\,|\,V_{qb}|^2\,m_b^5\over 192\pi^3} \,,
\end{equation}
and divide the differential decay rate into two pieces:
\begin{equation}\label{Gsplit}
   {1\over 2\Gamma_b}\,{\mbox{d}\Gamma\over\mbox{d}y}
   = F(y,\rho)\,\Theta(1-y-\rho) + S(y,\rho) \,.
\end{equation}
The first part is the result obtained in the free-quark decay model.
The function $F(y,\rho)$ is given by
\begin{equation}\label{Fyrho}
   F(y,\rho) = y^2\,\Big\{ 3 (1-\rho) (1-R^2) - 2 y (1-R^3) \Big\}
   \,;\quad R = {\rho\over 1-y} \,.
\end{equation}
In Fig.~\ref{fig:1}, we show $F(y,\rho)$ evaluated for $\rho=0.08$,
which is an appropriate value for $B\to X_c\,\ell\, \bar\nu$
transitions. The function $S(y,\rho)$ in (\ref{Gsplit}) contains the
nonperturbative corrections to a free-quark decay. The expression for
this functions obtained by naively constructing the operator product
expansion to next-to-leading order in $\Lambda/m_b$ is
\begin{eqnarray}\label{Sexpr}
   S(y,\rho) &=& y^2 \Bigg\{ {\lambda_1\over\big[ m_b(1-y)\big]^2}\,
    (3 R^2-4 R^3) \nonumber\\
   &&\mbox{}- {\lambda_1\over m_b^2(1-y)}\,(R^2-2 R^3)
    - {3\lambda_2\over m_b^2(1-y)}\,(2 R+3 R^2-5 R^3) \nonumber\\
   &&\mbox{}+ {\lambda_1\over 3 m_b^2}\,\Big[ 5y - 2(3-\rho) R^2
    + 4 R^3 \Big] \nonumber\\
   &&\mbox{}+ {\lambda_2\over m_b^2}\,\Big[ (6+5y) - 12 R
    - (9-5\rho) R^2 + 10 R^3 \Big] \Bigg\}
    \,\Theta(1-y-\rho) \nonumber\\
   &+& {\cal O}\Big[ \big(\Lambda/[m_b(1-y)]\big)^3 \Big] \,.
\end{eqnarray}
It is apparent that the operator product expansion gives an expansion
in two parameters, $\Lambda/m_b$ and $\Lambda/(m_b-2 E_\ell)=
\Lambda/[m_b(1-y)]$. Over most of the kinematic region, the terms in
$S(y,\rho)$ are of order $(\Lambda/m_b)^2$ or smaller. However,
provided that the parameter $\rho$ is of order $\Lambda/m_b$ or
smaller,\footnote{When $\rho$ is of order unity, $(1-y)$ cannot
become small in the physical region, and there is no problem with
(\ref{Sexpr}).}
the expansion becomes singular in the endpoint region: when $(1-y)$
is of order $\Lambda/m_b$, terms of order $\big(\Lambda/
[m_b(1-y)]\big)^n$ in $S(y,\rho)$ become of order unity. In
Fig.~\ref{fig:2}, we show $S(y,\rho)$ for $\rho=0.08$,
$\lambda_2=0.12$ GeV$^2$, and the two cases $\lambda_1=-0.1$ GeV$^2$
and $\lambda_1=-0.3$ GeV$^2$. The low-energy parameter $\lambda_2$
can be extracted from the known value of the $B^*$-$B$ mass
splitting. The average kinetic energy, and with it the value of
$\lambda_1$, are not well-known, however. Considering the various
theoretical arguments about this quantity that have been discussed in
the literature \cite{virial,Elet,BaBr}, we consider the two choices
given above as reasonable ``small'' and ``large'' values for
$\lambda_1$. Recently, Bigi et al.\ have derived the lower bound
$-\lambda_1\ge\frac{3}{2}\,\lambda_2$ in a quantum mechanical
framework \cite{Fermi}. However, since $\lambda_1$ and $\lambda_2$
have a different dependence on the renormalization point, this result
cannot be rigorous once short-distance corrections are taken into
account. In fact, it is possible to construct an explicit
counter-example to this bound using QCD sum rules \cite{Matt}.

The effect of nonperturbative corrections is very small over most of
the kinematic region. Close to the endpoint, however, a sharp spike
of height $\sim -\lambda_1/(m_b\rho)^2$ develops. The singular
behavior of the expansion becomes even more obvious when one goes to
higher orders; at order $1/m_b^3$, for instance, one encounters a
$\delta$-function at $y=1-\rho$. Clearly, one cannot believe the
shape of the function $S(y,\rho)$ in this region. To get a more
reliable description of the spectrum, the series in
$\Lambda/[m_b(1-y)]$ has to be resummed. In Ref.~\cite{shape}, it has
been shown that when the mass $m_q$ of the final-state quark is
neglected, such a resummation can be performed and leads to a smooth,
but rapidly varying, function. This so-called shape function is a
genuinely nonperturbative object, which can be defined in terms of
forward matrix elements of certain operators in the heavy quark
effective theory. A complete resummation of the operator product
expansion is, of course, a too complicated task. What can be achieved
is a summation of the leading terms in the limit $m_b\to\infty$ with
$m_b(1-y)$ kept fixed.

In this paper we shall elaborate on this proposal and extend it to
the case $m_q$ not being zero. Taking the limit $m_b\to\infty$ with
$m_b(1-y)$ fixed, we find from (\ref{Sexpr})
\begin{equation}\label{Slimit}
   S(y,\rho) \to {\lambda_1\,y^2\over\big[ m_b(1-y)\big]^2}\,
   (3 R^2-4 R^3)\,\Theta(1-y-\rho)
   + {\cal O}\Big[ \big(\Lambda/[m_b(1-y)]\big)^3 \Big] \,.
\end{equation}
Our goal will be to generalize this expression to all orders in
$\Lambda/[m_b(1-y)]$. In Sect.~\ref{sec:2}, we discuss in detail the
structure of the operator product expansion and explain the
significance of three different kinematic regions, which we will call
the free-quark decay region, the endpoint region, and the resonance
region. Our focus here is on the endpoint region, for which we
construct an appropriate approximation that resums the leading
singularities. We show that it is natural to distinguish the cases
where the parameter $\rho$ in (\ref{rhodef}) is of order
$\Lambda/m_b$ or $(\Lambda/m_b)^2$. We emphasize that there are, in
the context of the operator product expansion, incalculable
corrections to our results from bound-state effects in the final
state. In the two cases considered above, they are suppressed by a
factor $(\Lambda/m_b)^{1/2}$ or $\Lambda/m_b$, respectively. In
Sect.~\ref{sec:3}, we apply our formalism to inclusive semileptonic
$B$-meson decays and calculate the leading terms in the
charged-lepton energy spectrum. We show that, in both cases, the
spectrum can be written as a convolution of the free-quark decay rate
with a fundamental light-cone structure function. In
Sect.~\ref{sec:4}, we illustrate our results using a simple model,
which incorporates the known properties of the structure function and
provides a description of the decay spectrum in terms of a single
parameter. Sect.~\ref{sec:5} deals with a discussion of possibilities
to extract information about the structure function from experimental
data on the lepton spectra in inclusive semileptonic $B$-decays. In
Sect.~\ref{sec:6}, we summarize our results and give some
conclusions.

\section{Resummation of the Leading Endpoint Singularities}
\label{sec:2}

By the optical theorem, any inclusive decay rate can be related to
the imaginary part of a transition operator $T$, which is defined in
terms of the time-ordered product of two local operators. For the
cases at hand, this correlator is of the form
\begin{equation}\label{Tdefi}
   T(q^2,v\cdot q) = -i\int\mbox{d}^4 x\,e^{i(m_b v-q)\cdot x}\,
   \langle B(v)|\,T\,\Big\{ \bar b_v(x)\,\Gamma_1\,q(x),
   \bar q(0)\,\Gamma_2\,b_v(0) \Big\}\,|B(v)\rangle \,,
\end{equation}
where $q$ is the momentum transferred to the leptons, and $\Gamma_i$
are combinations of Dirac matrices. The $b$-quark field $b_v(x)$ is
related to the conventional field that appears in the QCD Lagrangian
by a phase redefinition:
\begin{equation}
   b_v(x) = \exp(-i m_b v\cdot x)\,b(x) \,.
\end{equation}
This is appropriate to make explicit a trivial but strong dependence
of the free quark field on the large mass $m_b$. Written in terms of
the rescaled fields, the hadronic matrix element in (\ref{Tdefi}) is
free of large momentum scales. Hence, the relevant momentum transfer
is
\begin{equation}
   Q = m_b v-q \,.
\end{equation}
Up to the difference between $m_b$ and $m_B$, the variable $Q^2$
corresponds to the invariant mass of the hadronic final state. As
long as $Q^2$ is large enough, it is legitimate to expand the
correlator in a tower of local operators of increasing dimension,
which are multiplied by coefficient functions that contain inverse
powers of $Q$. In general, this will be an expansion in three large
parameters: $m_b$, $v\cdot Q$, and $Q^2$. Over most of phase space,
$Q^2$ and $v\cdot Q$ scale with the heavy quark mass, and the
operator product expansion reduces to an expansion in powers of
$\Lambda/m_b$. There is, however, a kinematic region where $v\cdot Q$
is of order $m_b$, but $\Lambda^2\ll Q^2\ll m_b^2$. In this case, it
is consistent to work to leading order in $\Lambda/m_b$, but
necessary to keep higher-order terms in $Q^2$. In fact, as we shall
see, one has to resum these terms to all orders.

To leading order in $\Lambda/m_b$, the heavy quark field $b_v$ in
(\ref{Tdefi}) can be replaced by the corresponding two-component
spinor $h_v$ of the heavy quark effective theory \cite{Geor}.
Similarly, the physical $B$-meson states are replaced by the
corresponding states in the $m_b\to\infty$ limit. At tree-level, the
leading contribution to the correlator is obtained by contracting the
light-quark fields:
\begin{equation}
   T(q^2,v\cdot q) = \int\mbox{d}^4 x\,e^{i Q\cdot x}\,
   \langle B(v)|\,T\,\Big\{ \bar h_v(x)\,\Gamma_1\,S_q(x,0)\,
   \Gamma_2\,h_v(0) \Big\}\,|B(v)\rangle \,.
\end{equation}
Here $S_q(x,0)$ is the propagator of the $q$-quark in the background
field of the light constituents of the $B$-meson. The Fourier
transform of $S_q(x,0)$ has the form
\begin{eqnarray}\label{propag}
   S_q(Q) &=& {1\over\rlap{\,/}Q + i\rlap{\,/}D - m_q + i\epsilon}
    \nonumber\\
   &=& (\rlap{\,/}Q + i\rlap{\,/}D + m_q)\,{1\over
    (Q^2 + 2i Q\cdot D - m_q^2 - \rlap{\,/}D\,\rlap{\,/}D
    + i\epsilon)} \,.
\end{eqnarray}
Note that a derivative acting on the rescaled heavy quark field
corresponds to the residual momentum $k=p_b-m_b v$. Since the
residual momentum results from the interactions of the heavy quark
with light degrees of freedom, it is of order $\Lambda$.

There is a lot of information that can be deduced from the structure
of the background-field propagator. We will be interested in the
region where the components of $Q$ are large (of order $m_b$), since
otherwise the operator product expansion breaks down. To leading
order in $\Lambda/m_b$, we can then neglect the covariant derivative
in the numerator, as well as the term containing two derivatives in
the denominator. This gives
\begin{equation}\label{Sendp}
   S_q(Q) = {\rlap{\,/}Q + m_q\over Q^2 + 2i Q\cdot D - m_q^2
   + i\epsilon} + {\cal O}(\Lambda/m_b) \,.
\end{equation}
The term in the denominator containing the covariant derivative is of
order $\Lambda m_b$. To see whether it is important, we have to
distinguish different kinematic regions. When $Q^2\sim m_b^2$, the
derivative term is suppressed by a factor $\Lambda/m_b$. To leading
order, the propagator corresponds to the propagator of a free quark
with momentum $Q$. In this region, we thus recover the free-quark
decay model up to small nonperturbative corrections. However, in the
endpoint region, where $Q^2-m_q^2\sim\Lambda m_b$, it is not possible
to neglect the derivative term. This is why an expansion in powers of
$i Q\cdot D/Q^2$, which was applied in Refs.~\cite{Bigi,Blok,MaWe} to
derive (\ref{Sexpr}), becomes singular. Instead, one has to use
(\ref{Sendp}) for the propagator in the endpoint region. For even
smaller values $Q^2-m_q^2\sim \Lambda m_q$, the derivative term
becomes the dominant term in the denominator, and the operator
product expansion breaks down. This is the resonance region, where
the invariant mass $Q^2$ of the hadronic final state is close to the
mass-shell of the lowest-lying resonances containing a $q$-quark. The
fact that the resonance region is parametrically smaller than the
endpoint region implies that corrections to our results, which arise
from bound-state effects in the final state, are suppressed in the
$m_b\to\infty$ limit. This is fortunate, since these corrections are,
as a matter of principle, not calculable using an operator product
expansion.

Let us elaborate on this last point. To see to what level one is
sensitive to bound-state corrections in the final state, we replace
the quark mass $m_q$ by some effective mass $(m_q+\delta)$, with
$\delta$ of order $\Lambda$. The sensitivity to such effects depends
upon the size of $m_q$. For the hypothetical case $m_q\sim m_b$, the
corrections induced by $\delta\ne 0$ are of order $\Lambda m_b$ and
thus of the same magnitude as the term containing the covariant
derivative. In this case, the nonperturbative corrections to the
free-quark decay model are incalculable. In the opposite limit
$m_q\sim\Lambda$, corresponding to $\rho\sim (\Lambda/m_b)^2$, we
find that the effects corresponding to $\delta\ne 0$ are always
subleading. They are suppressed relative to the bound-state
corrections in the initial state by a factor $\Lambda/m_b$. This
shows that, in the analysis of the inclusive decays $B\to
X_u\,\ell\,\bar\nu$ and $B\to X_s\,\gamma$ \cite{shape,photon}, the
effect of a non-vanishing $u$- or $s$-quark mass, even as large as a
constituent mass, can be neglected. A very interesting case in
between the above is the situation where $m_q^2\sim\Lambda m_b$,
corresponding to $\rho\sim\Lambda/m_b$. We argue that this case
applies for $B\to X_c\,\ell\,\bar\nu$ decays, where $\rho\simeq 0.1$.
Then the term in the propagator containing the derivative is of the
same magnitude as $\rho$, and final-state corrections induced by
$\delta\ne 0$ are again subleading. However, they are only suppressed
by a factor $\Lambda/m_q\sim (\Lambda/m_b)^{1/2}$. We thus expect
that these corrections are more important than in the case of decays
into light (charmless) final states, and that the approach to the
$m_b\to\infty$ limit will be slower.

The above discussion shows that, for the relevant cases $\rho\sim
\Lambda/m_b$ and $\rho\sim(\Lambda/m_b)^2$, the propagator
(\ref{Sendp}) can be used to resum the leading terms in the endpoint
region. Corrections to it are suppressed in the $m_b\to\infty$ limit.
It is possible to simplify the result further. To this end, we write
$Q=(v\cdot Q)(n+\delta n)$ and choose the vectors $n$ and $\delta n$
such that $n$ is a null-vector on the forward light cone satisfying
$n^2=0$ and $n\cdot v=1$, and thus $v\cdot\delta n=0$. In the region
where the derivative term in (\ref{Sendp}) is important, i.e.\ for
$v\cdot Q$ of order $m_b$ but $Q^2$ of order $\Lambda m_b$, we have
$n\cdot\delta n\sim \Lambda/m_b$. As long as we restrict ourselves to
the leading term in the large-$m_b$ limit, we can thus neglect
$\delta n$. This leads to the final expression for the effective
background-field propagator:
\begin{equation}
   S_q(Q) \simeq \int\mbox{d}k_+\,\delta(k_+-i D_+)\,
   {\rlap{\,/}Q + m_q\over Q^2 + 2 k_+\,v\cdot Q - m_q^2
    + i\epsilon} \,,
\end{equation}
where we define $n\cdot D\equiv D_+$. In light-cone gauge, $n\cdot
A\equiv A_+=0$, the operator $i D_+$ reduces to $i\partial_+$ and
corresponds to the light-cone residual momentum $k_+$ of the heavy
quark inside the $b$-meson. Note that the appearance of the
null-vector $n$ is closely connected to our assumption that $m_q^2$
is of order $\Lambda m_b$ or $\Lambda^2$. For $m_q^2\sim m_b^2$, one
would have $Q^2\sim m_b^2$ even in the endpoint region, and $n$ would
be a time-like vector \cite{Fermi}. However, as we have argued above,
we do not believe that this case is of phenomenological importance.
{}From now on, we shall not consider it any more.

Let us use this form of the propagator to calculate the leading
contribution to the imaginary part of the correlator $T(q^2,q\cdot
v)$. We can simplify the Dirac structure of the hadronic matrix
element by using the identity
\begin{equation}
  \bar h_v\,\Gamma\,h_v = {1\over 2}\,\mbox{Tr}\,
  \big(\Gamma\,P_v\big)\,\bar h_v\,h_v - {1\over 2}\,
  \mbox{Tr}\,\big(\gamma_\mu\gamma_5\,P_v\,\Gamma\,P_v\big)\,
  \bar h_v\,\gamma^\mu\gamma_5\,h_v \,,
\end{equation}
which is valid for an arbitrary matrix $\Gamma$. Here
$P_v=\frac{1}{2}(1+\rlap/v)$ is a projector onto the large
components. Since the matrix element of the axial vector current
between $B$-meson states vanishes by parity invariance, we obtain for
the leading term
\begin{eqnarray}\label{Tmaster}
   {1\over\pi}\,\mbox{Im}\,T(q^2,v\cdot q) &=& - {1\over 4}\,
    \mbox{Tr}\,\Big\{ \Gamma_1\,(\rlap{\,/}Q + m_q)\,\Gamma_2\,
    (1+\rlap/v) \Big\} \nonumber\\
   &&\times\int\mbox{d}k_+\,f(k_+)\,
    \delta(Q^2 + 2 k_+\,v\cdot Q - m_q^2) \,,
\end{eqnarray}
where
\begin{equation}\label{fdef}
   f(k_+) = \langle B(v)|\,\bar h_v\,\delta(k_+-i D_+)\,h_v\,
   |B(v)\rangle
\end{equation}
is the leading-twist structure function that determines the
probability to find a $b$-quark with light-cone residual momentum
$k_+$ inside the $B$-meson \cite{photon}. We use a mass-independent
normalization of states such that $\langle B(v)|\,\bar h_v\,h_v\,
|B(v)\rangle=1$. It then follows that the structure function is
normalized to unity:
\begin{equation}\label{fnorm}
   \int\mbox{d}k_+\,f(k_+) = 1 \,.
\end{equation}
It will later be convenient to introduce the Fourier transform of the
structure function, which is given by the forward matrix element of a
bilocal operator:\footnote{The definition of a heavy quark structure
function as the Fourier transform of the bilocal operator in
(\ref{bilocal}) was used, in a different context, in
Ref.~\cite{Jaffe}.}
\begin{eqnarray}\label{bilocal}
   \widetilde f(t) &=& \int\mbox{d}k_+\,e^{-i k_+ t}\,f(k_+)
    \nonumber\\
   &=& \langle B(v)|\,\bar h_v(0)\,P\exp\bigg[ -i\int\limits_0^t\!
    \mbox{d}u\,A_+(u n)\bigg]\,h_v(t n)\,|B(v)\rangle \,.
\end{eqnarray}
This function obeys the normalization condition $\widetilde f(0)=1$.
Note that the path-ordered exponential is absent in light-cone gauge.

Let us recall at this point some important properties of the
structure function \cite{shape,photon}. The moments of $f(k_+)$ are
given by forward matrix elements of leading-twist, higher-dimension
operators in the heavy quark effective theory. They form a set of
low-energy parameters $A_n$ defined by
\begin{equation}\label{Anrela}
   A_n = \int\mbox{d}k_+\,k_+^n\,f(k_+) = i^n\,\widetilde f^{(n)}(0)
   = \langle B(v)|\,\bar h_v\,(i D_+)^n\,h_v\,|B(v)\rangle \,,
\end{equation}
where $\widetilde f^{(n)}(0)$ is a short-hand notation for the $n$-th
derivative of $\widetilde f(t)$ evaluated at $t=0$. Using the
equation of motion of the heavy quark effective theory, one obtains
for the first three moments
\begin{eqnarray}\label{moments}
   A_1 &=& 0 \,,\qquad A_2 = -{\lambda_1\over 3} \,, \nonumber\\
   A_3 &=& -{v_\nu\over 6}\,\langle B(v)|\,\bar h_v\,\big[
    D_\mu,g_s G^{\mu\nu}\big]\,h_v\,|B(v)\rangle \,,
\end{eqnarray}
where $g_s G^{\mu\nu}=i[D^\mu,D^\nu]$ is the gluon field-strength
tensor, and $\lambda_1$ has been introduced in (\ref{lamdef}).
Assuming that $0.1~\mbox{GeV}^2<-\lambda_1< 0.3~\mbox{GeV}^2$, we
find that $180~\mbox{MeV}<\sqrt{A_2}< 315~\mbox{MeV}$. This quantity
is related to the characteristic width of the endpoint region
\cite{shape,photon}. Using the equation of motion for the gluon
field, the moment $A_3$ may be written in terms of matrix elements of
four-quark operators, which can be evaluated in the factorization
approximation \cite{Fermi,Thomnew}. This leads to the rough estimate
\begin{equation}
   A_3\approx -{2\pi\over 27}\,\alpha_s\,f_B^2\,m_B
   \approx -(270~\mbox{MeV})^3 \,,
\end{equation}
where we have assumed $f_B\approx 200$ MeV and $\alpha_s\approx 0.4$.
Summarizing these results, we know that the light-cone structure
function is centered around $k_+=0$, has a width of order 200-300 MeV
determined by the average kinetic energy of the $b$-quark inside the
$B$-meson, and most likely (if factorization holds approximately) has
an asymmetry towards negative values of $k_+$.

The support of the structure function can be deduced by observing
that the total light-cone momentum fraction
\begin{equation}
   x = {(p_b)_+\over(p_B)_+} = {m_b+k_+\over m_B}
\end{equation}
must be bounded between 0 and 1. It follows that $-m_b\le k_+\le
m_B-m_b$. Since the structure function is defined in the heavy quark
effective theory, corresponding to the $m_b\to\infty$ limit, this
implies that
\begin{equation}\label{krange}
   -\infty < k_+ \le \bar\Lambda \,,
\end{equation}
where $\bar\Lambda$ denotes the asymptotic value of the mass
difference between a heavy meson and the heavy quark that it
contains, and can be identified with the effective mass of the light
degrees of freedom interacting with the heavy quark \cite{AMM}. We
expect that the support of $f(k_+)$ for negative values of $k_+$ is
slightly larger (because of the asymmetry) but of the same order of
magnitude, i.e.\ $f(k_+)$ should be exponentially small for $k_+\ll
-\bar\Lambda$.

\section{Calculation of the Lepton Spectrum}
\label{sec:3}

Let us now proceed to calculate the charged-lepton spectrum in
semileptonic $B$-decays. The matrices $\Gamma_i$ in (\ref{Tmaster})
are of the form $\gamma^\mu(1-\gamma_5)$, and we obtain for the
leading contribution to the correlator
\begin{eqnarray}
   {1\over\pi}\,\mbox{Im}\,T^{\mu\nu}(q^2,v\cdot q)
   &=& -\int\mbox{d}k_+\,f(k_+)\,
    \delta(Q^2 + 2 k_+\,v\cdot Q - m_q^2) \nonumber\\
   &&\times \Big[ Q^\mu v^\nu + Q^\nu v^\mu - g^{\mu\nu} v\cdot Q
    - i\epsilon^{\mu\nu\alpha\beta} Q_\alpha v_\beta \Big] \,.
\end{eqnarray}
The next step is to express $Q$ in terms of the lepton momentum $q$,
and to contract $T^{\mu\nu}$ with the leptonic tensor. A
straightforward calculation leads to the triple-differential decay
rate
\begin{eqnarray}
   {\mbox{d}^3\Gamma(B\to X_q\,\ell\,\bar\nu)\over
    \mbox{d}q^2\,\mbox{d}(v\cdot q)\,\mbox{d}E_\ell}
   &=& {G_F^2\,|\,V_{qb}|^2\over 2\pi^3}\,(v\cdot q-E_\ell)\,
    (2 m_b E_\ell - q^2) \int\mbox{d}k_+\,f(k_+) \nonumber\\
   &&\times \delta\Big[ q^2 - 2(m_b+k_+)\,v\cdot q + m_b^2
    + 2 k_+ m_b - m_q^2 \Big] \,,\qquad\mbox{}
\end{eqnarray}
where $E_\ell=v\cdot p_\ell$ denotes the charged-lepton energy in the
parent rest frame. To the order we are working, i.e.\ to leading
order in the large-$m_b$ limit, we can rewrite this expression in
such a way that all dependence on $m_b$ and $k_+$ comes through the
combination
\begin{equation}
   m_b^*(k_+) = m_b + k_+ \,,
\end{equation}
which we shall identify with the effective mass of the $b$-quark
inside the $B$-meson. We thus observe that the leading bound-state
corrections amount to averaging the parton-model rate for the decay
of a quark with mass $m_b^*(k_+)$ over the distribution function
$f(k_+)$. The free-quark decay model is recovered in the limit
$f(k_+)\to\delta(k_+)$.

Next we integrate over $v\cdot q$, and then over $q^2$ in the
kinematic region
\begin{equation}
   0 \le q^2 \le 2 E_\ell\,m_b^*\,(1-R_*) \,;\quad
   R_* = {m_q^2\over m_b^*(m_b^* - 2 E_\ell)} \,,
\end{equation}
which follows from the requirement that $0\le q^2\le 4 E_\ell E_\nu$.
We obtain
\begin{eqnarray}\label{convol}
   {\mbox{d}\Gamma\over\mbox{d}E_\ell}
   &=& {G_F^2\,|\,V_{qb}|^2\over 12\pi^3}\,E_\ell^2
    \int\mbox{d}k_+\,f(k_+)\,\Theta\Big[ m_b^{*2} - m_q^2
    - 2 m_b^*\,E_\ell \Big] \nonumber\\
   &&\times \Big\{ 3 (m_b^{*2} - m_q^2)(1 - R_*^2)
    - 4 m_b^*\,E_\ell\,(1 - R_*^3) \Big\} \,.
\end{eqnarray}
This result has several interesting properties. We first note that
the heavy quark mass $m_b$ does no longer appear explicitly. For
this reason, and in particular when the focus is on the endpoint
region, it would be unnatural to introduce the rescaled lepton energy
$y=2 E_\ell/m_b$. Hence, we will hereafter present our results as
function of the lepton energy $E_\ell$, which is the quantity that is
actually measured in experiments. Note, in particular, that (to the
order we are working) the maximum value of the lepton energy is
correctly reproduced. From the fact that $k_+^{\rm max}=
\bar\Lambda=m_B-m_b$ according to (\ref{krange}), it follows that
\begin{equation}\label{Elimit}
   (m_b^*)_{\rm max} = m_B \,,\qquad
   E_\ell^{\rm max} = {m_B\over 2}\,\bigg( 1
   - {m_q^2\over m_B^2}\bigg) \,.
\end{equation}
This should be compared to the kinematic endpoint $E_\ell^{\rm max}
= (m_b/2)(1-\rho)$ predicted by the free-quark decay model. The
difference between the heavy quark mass and the physcial $B$-meson
mass is correctly accounted for in our approach. Note, however, that
we are not able to account for the fact that, instead of the quark
mass $m_q$, there should appear in (\ref{Elimit}) the mass of the
lightest meson containing the $q$-quark. As discussed in
Sect.~\ref{sec:2}, this effect is subleading, i.e.~it vanishes in the
$m_b\to\infty$ limit, whereas the difference between $m_B$ and $m_b$
remains.

For lepton energies not too close to the endpoint, the difference
between $m_b$ and the effective mass $m_b^*$ becomes irrelevant, and
(\ref{convol}) reduces to the result of the free-quark decay model,
as given by the first term in (\ref{Gsplit}). Since the first moment
$A_1$ in (\ref{moments}) vanishes, the nonperturbative corrections in
this region are of order $(\Lambda/m_b)^2$. Note, in particular, that
\begin{equation}
   \int\mbox{d}k_+\,f(k_+)\,\big[ m_b^*(k_+)\big]^n
   = m_b^n\,\bigg\{ 1 - {n(n-1)\over 6}\,{\lambda_1\over m_b^2}
   + \ldots \bigg\} \,,
\end{equation}
meaning that, up to second-order corrections, it is in fact the mass
of the $b$-quark that matters in the main region of phase space
\cite{Chay}.

It is clear from this discussion that substantial nonperturbative
corrections show up only in the endpoint region, where $(m_b-2
E_\ell)$ is of order $\Lambda$, and the difference between $m_b$ and
$m_b^*$ becomes important. We can separate these effects from the
free-quark decay distribution by subtracting a term $\delta(k_+)$
from the structure function. In this way, we obtain the lepton
spectrum as a sum of two terms, as shown in (\ref{Gsplit}).
Introducing then the rescaled lepton energy $y=2 E_\ell/m_b$ and the
mass ratio $\rho=m_q^2/m_b^2$, and keeping only the leading
contributions in the large-$m_b$ limit, we obtain for the shape
function defined in (\ref{Gsplit})
\begin{equation}
   S(y,\rho) = y^2\!\!\!\int\limits_{m_b(y-1+\rho)}^{\bar\Lambda}
    \!\!\!\mbox{d}k_+\,\big[ f(k_+) - \delta(k_+) \big]\,
    (1 - 3 R_*^2 + 2 R_*^3) +~\mbox{less singular terms,}
\end{equation}
where now
\begin{equation}
   R_* = {m_b \rho\over m_b(1-y) + k_+} \,.
\end{equation}
This is the correct generalization of (\ref{Slimit}). If we would
formally expand the structure function (\ref{fdef}) as
\begin{equation}\label{fexpand}
   f(k_+) = \delta(k_+) + \sum_{n=2}^\infty\,{(-1)^n\over n!}\,
   A_n\,\delta^{(n)}(k_+)
\end{equation}
and truncate the series at $n=2$, we would recover the singular term
shown in (\ref{Slimit}). Such a truncation is not justified, however,
since every term in (\ref{fexpand}) is of the same magnitude.

For completeness, we note that the form of the convolution
(\ref{convol}) simplifies in the limit $m_q\to 0$, which is relevant
for $B\to X_u\,\ell\,\bar\nu$ decays. One obtains
\begin{eqnarray}\label{buconv}
   {\mbox{d}\Gamma\over\mbox{d}E_\ell}
   &=& {G_F^2\,|\,V_{qb}|^2\over 12\pi^3}\,E_\ell^2
    \int\mbox{d}k_+\,f(k_+)\,\Theta(m_b^* - 2 E_\ell)\,
    m_b^*\,(3 m_b^* - 4 E_\ell) \nonumber\\
   &\simeq& {G_F^2\,|\,V_{qb}|^2\over 12\pi^3}\,m_b\,E_\ell^2\,
    (3 m_b - 4 E_\ell)\!\!\!\int\limits_{2 E_\ell-m_b}^{\bar\Lambda}
    \!\!\!\mbox{d}k_+\,f(k_+) \,.
\end{eqnarray}
This agrees with the result obtained in
Refs.~\cite{shape,photon,Fermi}.

\section{A Realistic Model}
\label{sec:4}

\begin{figure}[h]
   \vspace{0.5cm}
   \epsfysize=7cm
   \centerline{\epsffile{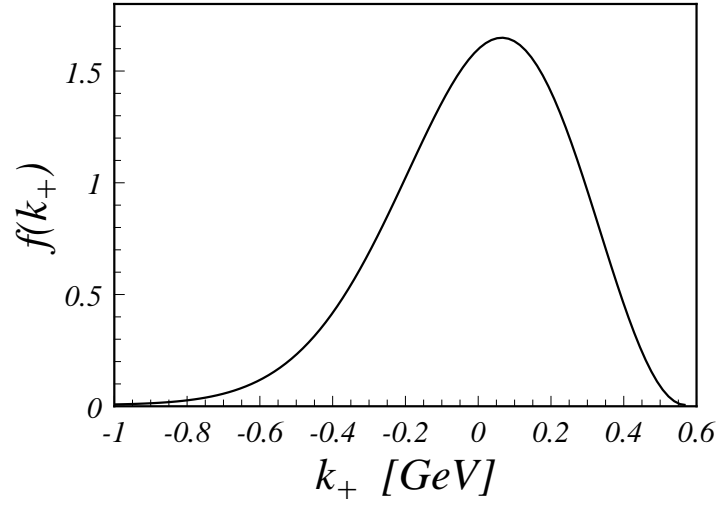}}
   \centerline{\parbox{11cm}{\caption{\label{fig:3}
Model ansatz (\protect\ref{ftoy}) for the structure function
$f(k_+)$, evaluated for $\bar\Lambda=0.57$ GeV.
   }}}
\end{figure}

At this point, it is instructive to illustrate our results with a
simple, but realistic model. To this end, we propose the following
one-parameter ansatz for the light-cone structure function:
\begin{equation}\label{ftoy}
   f(k_+) = {32\over\pi^2\bar\Lambda}\,(1-x)^2\,
   \exp\bigg\{ - {4\over\pi}\,(1-x)^2 \bigg\}\,
   \Theta(1-x) \,;\quad x = {k_+\over\bar\Lambda} \,,
\end{equation}
where $\bar\Lambda=m_B-m_b$ is treated as a free parameter. Below we
shall use the value $\bar\Lambda=0.57$ GeV, which is predicted by QCD
sum rules \cite{Baga,sumrul}. Our model structure function is shown
in Fig.~\ref{fig:3}. It obeys all requirements that have been pointed
out in Sect.~\ref{sec:2}. The support of $f(k_+)$ is limited to
values $k_+<\bar\Lambda$. The integral over the structure function is
normalized to unity, and the first moment vanishes. The higher
moments are proportional to powers of $\bar\Lambda$. In particular,
we obtain
\begin{eqnarray}
   A_2 &=& - {\lambda_1\over 3}
    = \bigg( {3\pi\over 8} - 1 \bigg)\,\bar\Lambda^2
    \simeq (0.42\,\bar\Lambda)^2 \,, \nonumber\\
   A_3 &=& - \bigg( 2 - {5\pi\over 8} \bigg)\,
    \bar\Lambda^3\simeq - (0.33\,\bar\Lambda)^3 \,.
\end{eqnarray}
For $\bar\Lambda=0.57$ GeV, we find $\lambda_1\simeq -0.17$ GeV$^2$
and $A_3\simeq -(190~\mbox{MeV})^3$. These numbers agree well with
our estimates in Sect.~\ref{sec:2}. Given these low-energy
parameters, we can compute the $b$-quark mass from the expansion
\cite{FaNe}
\begin{equation}
   m_B = m_b + \bar\Lambda - {\lambda_1+3\lambda_2\over 2 m_b^2}
   + \ldots \,.
\end{equation}
In our model, we obtain $m_b\simeq 4.71$ GeV. Similarly, we find
$m_c\simeq 1.35$ GeV for the charm-quark mass, and hence $\rho\simeq
0.08$. We shall use this set of parameters throughout the paper.

\begin{figure}[h]
   \vspace{0.5cm}
   \epsfysize=7cm
   \centerline{\epsffile{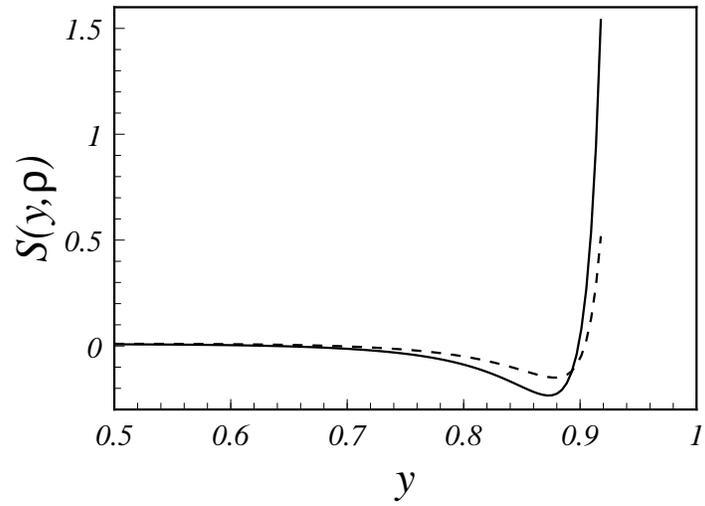}}
   \centerline{\parbox{11cm}{\caption{\label{fig:2}
Nonperturbative corrections to the lepton spectrum obtained using a
naive operator product expansion. The solid line corresponds to
$\lambda_1=-0.3$ GeV$^2$, the dashed one to $\lambda_1=-0.1$
GeV$^2$.
   }}}
\end{figure}

\begin{figure}[h]
   \vspace{0.5cm}
   \epsfysize=7cm
   \centerline{\epsffile{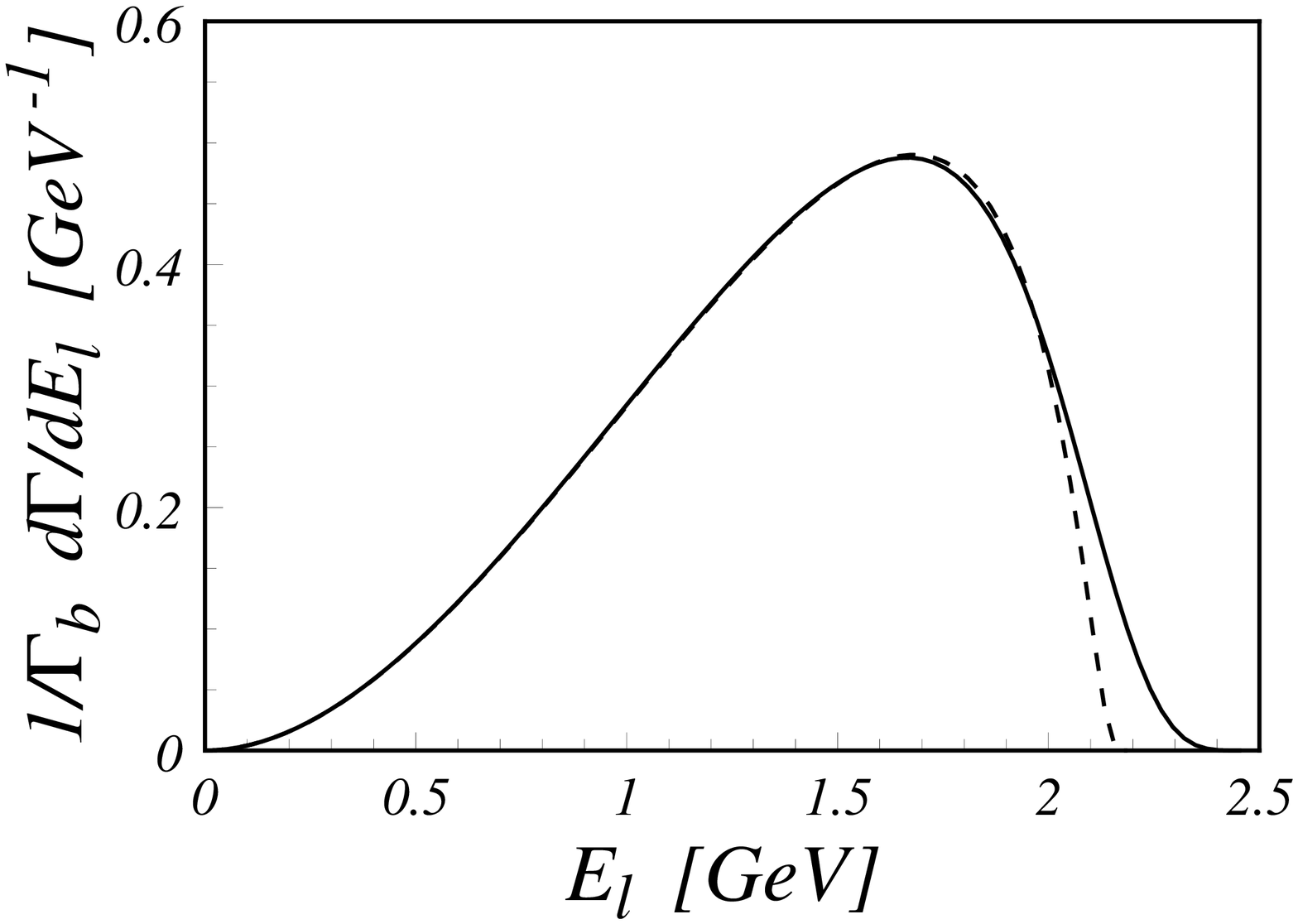}}
   \vspace{0.5cm}
   \epsfysize=7cm
   \centerline{\epsffile{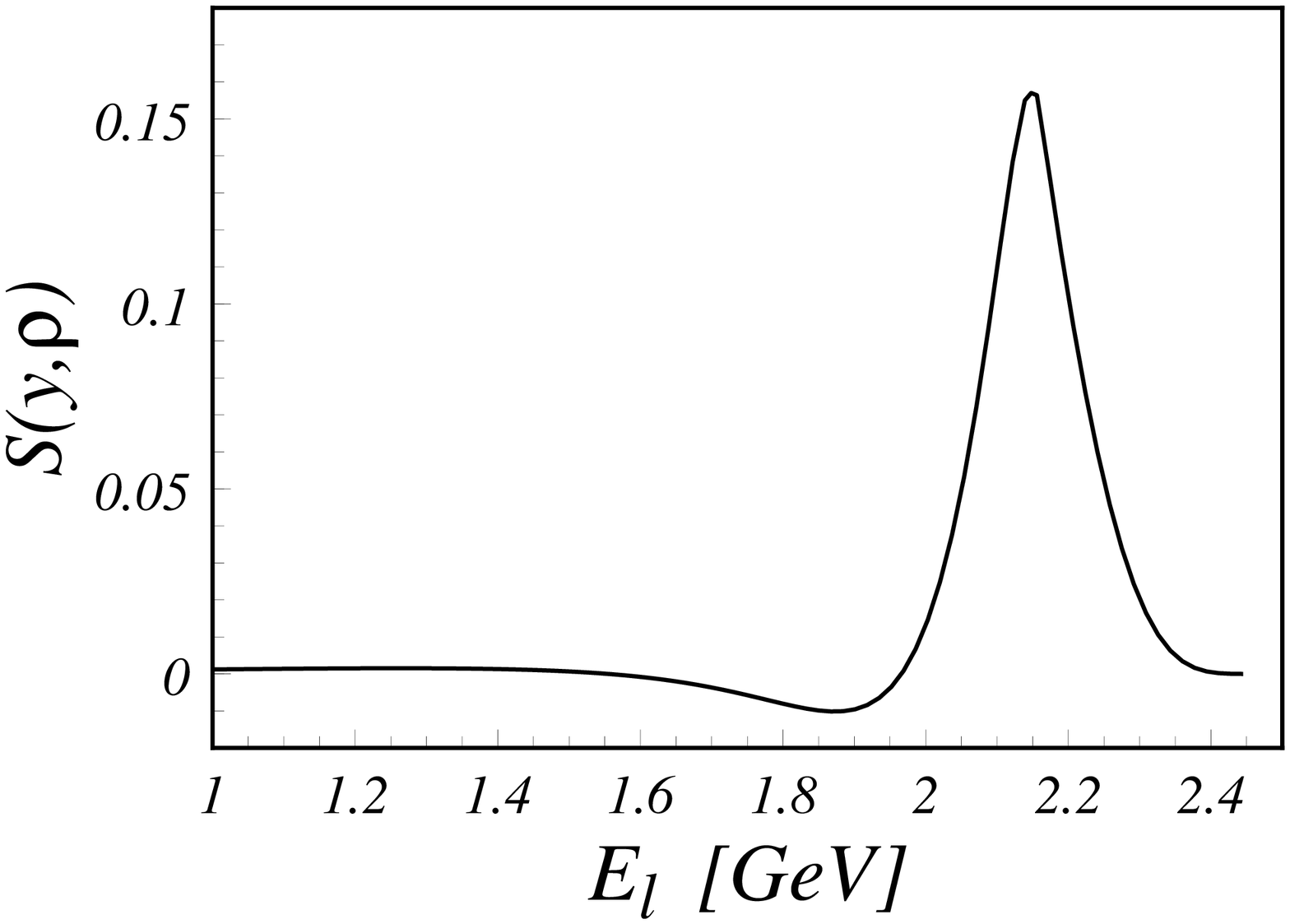}}
   \centerline{\parbox{11cm}{\caption{\label{fig:4}
(a) Charged-lepton spectrum $(1/\Gamma_b)\,\mbox{d}\Gamma/
\mbox{d}E_\ell$ in $B\to X_c\,\ell\,\bar\nu$ decays. The solid line
is obtained from the convolution in (\protect\ref{convol}) using the
ansatz (\protect\ref{ftoy}) for the structure function. The dashed
line shows the prediction of the free-quark decay model. (b) The
shape function $S(y,\rho)$, which is obtained from the difference of
the two curves in (a).
   }}}
\end{figure}

In Fig.~\ref{fig:4}a, we show the lepton spectrum $(1/\Gamma_b)\,
\mbox{d}\Gamma/\mbox{d}E_\ell$ obtained from (\ref{convol}) in
comparison with the free-quark decay distribution. The difference
between the two spectra is shown in Fig.~\ref{fig:4}b. It corresponds
to the shape function, which describes the leading nonperturbative
corrections to the free-quark decay model. For the ease of comparison
with Fig.~\ref{fig:2}, we have multiplied the vertical scale by
$m_b/4$ in order to comply with the definition of $S(y,\rho)$ in
(\ref{Gsplit}). In order to convert the horizontal scale to the
variable $y$, one would have to multiply $E_\ell$ by $2/m_b$. We find
that the shape function is indeed sizable only in the endpoint
region. Comparing the result of our resummation to the singular form
of $S(y,\rho)$ given in (\ref{Sexpr}) and shown in Fig.~\ref{fig:2},
we see that the resummation has eliminated the unrealistic spike at
the endpoint. What remains is a smooth function, which is rapidly
varying on scales of order $\Delta y\sim 1$. In other words, the
decay probabilities have been redistributed such that the height of
$S(y,\rho)$ is strongly reduced; however, and most importantly, the
shape function now extends beyond the parton-model endpoint (which
would correspond to 2.17 GeV). Needless to say, the precise shape of
$S(y,\rho)$ depends on the form of the structure function $f(k_+)$.
In the following section, we shall discuss some strategies how to
extract the structure function from data.

\begin{figure}[h]
   \vspace{0.5cm}
   \epsfysize=7cm
   \centerline{\epsffile{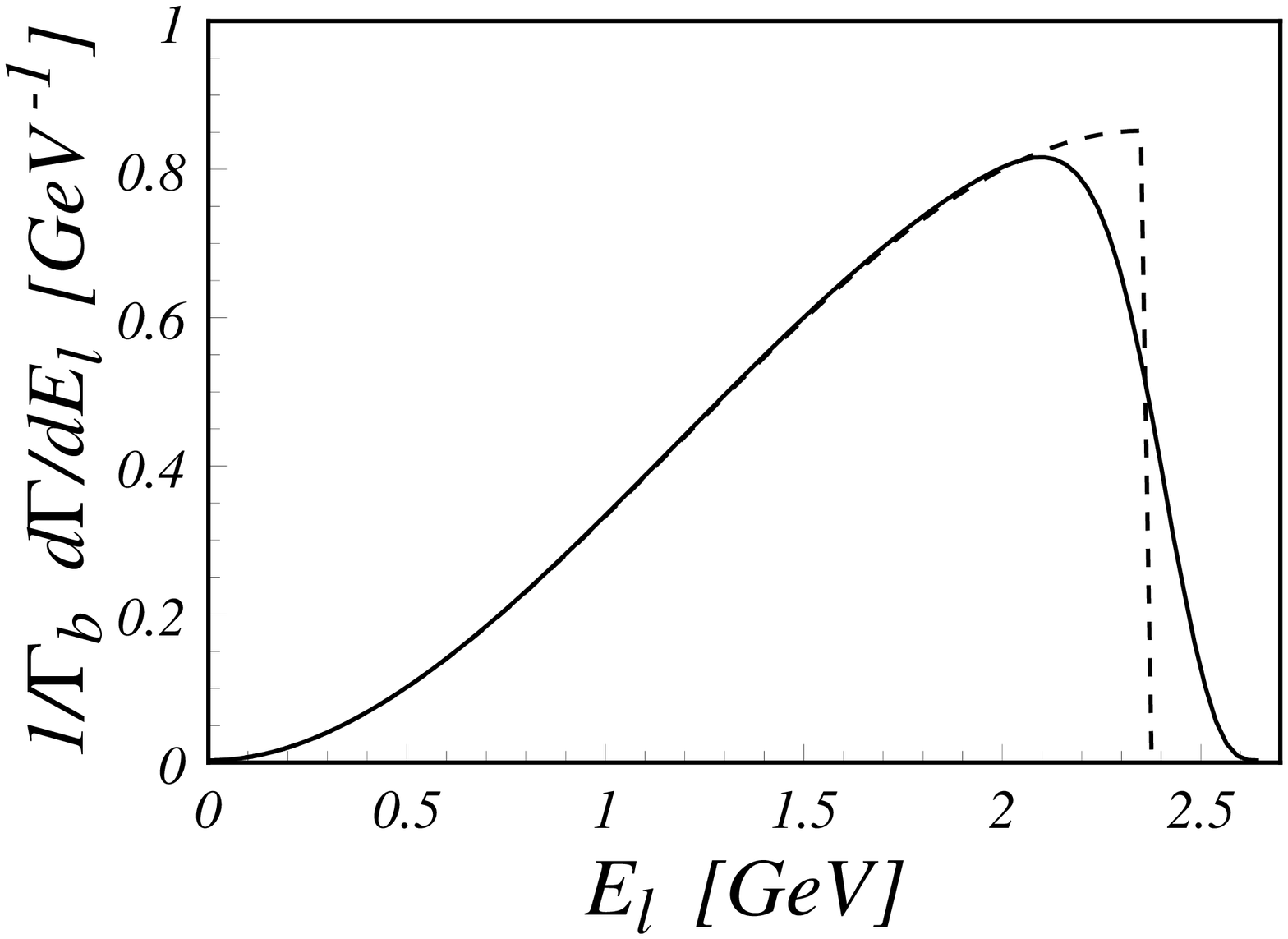}}
   \vspace{0.5cm}
   \epsfysize=7cm
   \centerline{\epsffile{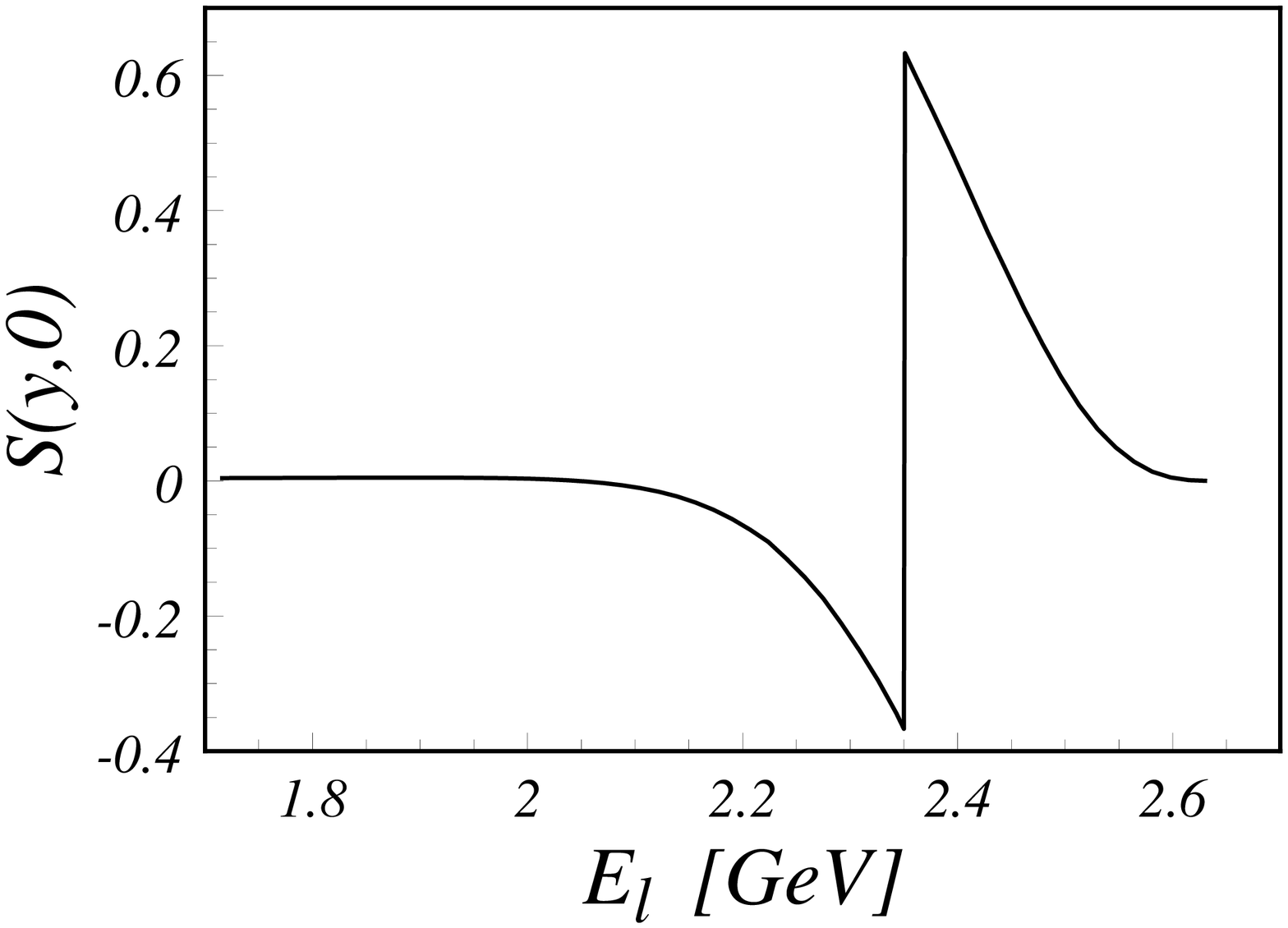}}
   \centerline{\parbox{11cm}{\caption{\label{fig:5}
Same as Fig.~\protect\ref{fig:4}, but for $B\to X_u\,\ell\,\bar\nu$
decays.
   }}}
\end{figure}

In Fig.~\ref{fig:5}, we show the corresponding spectra for charmless
$B\to X_u\,\ell\,\bar\nu$ decays, setting $\rho=0$. The
nonperturbative effects in the endpoint region become more
pronounced, due to the fact that the free-quark decay distribution
ends with a step-function in this case.

\section{Extraction of the Structure Function}
\label{sec:5}

In this section, we briefly discuss how one could, in principle,
extract information about the structure function $f(k_+)$ from
experimental data on inclusive semileptonic $B$-decays. We should
mention from the beginning that our discussion will be incomplete in
that it neglects radiative corrections as well as the effects of
experimental uncertainties. Both could be important. It is also clear
from the results of Refs.~\cite{photon,Fermi} that a much better
place to extract the structure function would be the photon spectrum
in inclusive $B\to X_s\,\gamma$ decays. However, in view of the fact
that this spectrum will be very hard to measure in the near future,
whereas very detailed data for semileptonic decay spectra already
exist \cite{CLEO}, we think it is worthwhile to consider
possibilities to extract at least some relevant information from
semileptonic decays as well.

We start with a discussion of the charmless decays $B\to
X_u\,\ell\,\bar\nu$. They are particularly simple, since the
nonperturbative corrections in (\ref{buconv}) are contained in an
integral over the structure function. Up to an overall normalization
factor (which depends on $V_{cb}$ and $m_b$), one can directly
extract the function $F(E_\ell)$ defined as
\begin{equation}
   F(E_\ell) = \!\!\!\int\limits_{2 E_\ell-m_b}^{\bar\Lambda}
   \!\!\!\mbox{d}k_+\,f(k_+) \propto
   {1\over E_\ell^2\,(3 m_b-4 E_\ell)}\,
   {\mbox{d}\Gamma\over\mbox{d}E_\ell}
\end{equation}
from a measurement of the lepton spectrum. The normalization can be
fixed by observing that $F(E_\ell)$ must approach unity when $m_b-2
E_\ell\gg\Lambda$. Using the definition (\ref{fdef}) of the structure
function, we obtain
\begin{equation}
   F(E_\ell) = \langle B(v)|\,\bar h_v\,
   \Theta(m_b - 2 E_\ell + i D_+)\,h_v\,|B(v)\rangle \,.
\end{equation}
The derivative of $F(E_\ell)$ with respect to the lepton energy gives
the structure function evaluated at $k_+=2 E_\ell-m_b$:
\begin{equation}
   F'(E_\ell) = - 2\,f(2 E_\ell-m_b) \,.
\end{equation}
Given a measurement of $F(E_\ell)$, one can extract the moments $A_n$
of the structure function, which have been defined in (\ref{Anrela}),
by integration with appropriate weight functions. One finds that
\cite{shape}
\begin{equation}
   \int\mbox{d}E_\ell\,(2 E_\ell-m_b)^n\,\Big[ F(E_\ell)
   - \Theta(2 m_b-E_\ell) \Big] = {A_{n+1}\over 2(n+1)} \,.
\end{equation}
The fact that $A_1=0$ can be used to fix the value of $m_b$ in the
step-function. In practice, the presence of several sources of
experimental and theoretical uncertainties (in particular, radiative
corrections and corrections of order $\Lambda/m_b$, which we neglect
in the above expressions) will probably limit this extraction method
to the first few moments.

Let us now turn to the case of $B\to X_c\,\ell\,\bar\nu$ decays,
where the form of the convolution integral (\ref{convol}) is more
complicated. Let us assume that the parameters $m_b$ and $m_c$ have
been extracted from a fit of the spectrum to the free-quark decay
model in the region far away from the endpoint. One can then write
the observed lepton spectrum in terms of the parameter
\begin{equation}
   \varepsilon = 2 E_\ell - m_b \,.
\end{equation}
To leading order in the large-$m_b$ limit, the differential decay
rate (\ref{convol}) can be rewritten in the form of a convolution of
the structure function $f(k_+)$ with the parton-model distribution
function $G(\varepsilon)$,
\begin{equation}
   {1\over E_\ell^2}\,{\mbox{d}\Gamma\over\mbox{d}E_\ell}
   \equiv F(\varepsilon)
   = \int\mbox{d}k_+\,f(k_+)\,G(\varepsilon-k_+) \,,
\end{equation}
where, according to (\ref{Fyrho}),
\begin{equation}
   G(\varepsilon) \propto \Theta(-m_b\rho-\varepsilon)\,
   \Big\{ 3 m_b(1-\rho)(1-R^2) - 2(m_b+\varepsilon)(1-R^3) \Big\}
\end{equation}
up to an overall factor, and $R=-m_b\rho/\varepsilon$. Let us now
introduce the Fourier transforms
\begin{eqnarray}
   \widetilde F(t) &=& \int\mbox{d}\varepsilon\,
    e^{-i\varepsilon t}\,F(\varepsilon) \,, \nonumber\\
   \widetilde G(t) &=& \int\mbox{d}\varepsilon\,
    e^{-i\varepsilon t}\,G(\varepsilon) \,.
\end{eqnarray}
Using (\ref{bilocal}), we find that
\begin{equation}
   \widetilde f(t) = {\widetilde F(t)\over\widetilde G(t)} \,,
\end{equation}
i.e., by taking the ratio of the Fourier transforms of the observed
lepton spectrum and of the parton-model distribution, one can in
principle extract the matrix element of the bilocal operator in
(\ref{bilocal}). The Fourier transform of $\widetilde f(t)$ gives the
structure function. Note that the normalization of the spectra is
irrelevant in this context, since we know the normalization of
$\widetilde f(t)$ at the origin: $\widetilde f(0)=1$.

Given an experimental determination of $\widetilde F(t)$, one can in
principle compute the low-energy parameters $A_n$ using
(\ref{Anrela}), i.e.
\begin{equation}
   A_n = \bigg(i\,{\partial\over\partial t}\bigg)^n
   \Bigg({\widetilde F(t)\over\widetilde G(t)}\Bigg)\Bigg|_{t=0} \,.
\end{equation}
In practice, such an analysis will again most likely be limited to
the first few moments.

A less ambitious approach that could be taken is to rely on a
theory-inspired ansatz for the structure function $f(k_+)$, which
should depend upon few parameters with a well-defined physical
meaning. The goal would be to extract these parameters from a fit to
experimental data. This procedure is familiar from the present
analysis of inclusive decay spectra in the context of the
phenomenological model of Altarelli et al.~\cite{ACM}. We believe
that a reasonable parameterization of $f(k_+)$ should contain (i) the
parameter $\bar\Lambda$, which determines the gap between the
parton-model endpoint and the physical endpoint of the lepton
spectrum, (ii) the parameter $\lambda_1$, which is proportional to
the width of the endpoint region, and (iii) an asymmetry parameter,
which is related to the third moment $A_3$ of the distribution
function. An extraction of these fundamental quantities, even if it
is affected by substantial uncertainties, would be most desirable.

\section{Conclusions}
\label{sec:6}

In the endpoint region of the lepton spectrum in inclusive
semileptonic decays of $B$-mesons, a naive operator product expansion
in powers of $\Lambda/m_b$ breaks down. The reason is that in this
region of phase space, there emerges a second expansion parameter,
$\Lambda/(2 m_b-E_\ell)$, which is much larger than $\Lambda/m_b$. In
this case, a partial resummation of the operator product expansion is
necessary before the theoretical results can be compared with data.
This resummation is such that one sums all contributions of order
$[\Lambda/(2 m_b-E_\ell)]^n$, keeping however only the leading terms
in $\Lambda/m_b$. In many respects, it resembles the summation of
leading-twist contributions in deep-inelastic scattering.

For the cases of $B\to X_u\,\ell\,\bar\nu$ and $B\to X_s\,\gamma$
decays, where the mass of the quark in the final state can be
neglected, it has been shown in Refs.~\cite{shape,photon,Fermi} that
the leading nonperturbative contributions close to the endpoint may
be resummed into a light-cone structure function $f(k_+)$, which
gives the probability to find a $b$-quark with light-cone residual
momentum $k_+$ inside the $B$-meson. This function is defined in
terms of forward matrix elements in the heavy quark effective theory.
It is thus independent of $m_b$ and has a universal character. In the
present paper, we have shown that the same structure function
describes the inclusive decays in the presence of a finite quark-mass
in the final state, provided that the ratio $\rho=(m_q/m_b)^2$ is
consistently treated as being of order $\Lambda/m_b$. In this case,
however, the nonperturbative corrections to the leading behavior are
of order $(\Lambda/m_b)^{1/2}\sim \Lambda/m_q$. Some of these
corrections are due to hadronization effects in the final state and
appear to be incalculable in the context of the operator product
expansion. Numerically, they could be quite significant. Yet, since
$\rho\simeq 0.08$ for case of $B\to X_c\,\ell\,\bar\nu$ decays, we
argue that our results should apply and should provide a correct
description of the leading nonperturbative effects. We should point
out that a different position was taken by Bigi et al.~\cite{Fermi},
where it was argued that $b\to c$ transitions are close to the
so-called ``small velocity limit'', where $m_b-m_c$ is assumed to be
much smaller than $m_b$, corresponding to $\rho\sim 1$. In this case,
even the leading effects are no longer described by the universal
light-cone distribution function. Which of the two cases is closer to
reality remains to be seen. We note, however, that the experimental
fact that semileptonic $B$-decays are not saturated by the exclusive
modes $B\to D^{(*)}\ell\,\bar\nu$ indicates that the ``small velocity
limit'', in which these two modes would saturate the total rate, can
only be a rather crude approximation.

In both cases, $B\to X_u\,\ell\,\bar\nu$ and $B\to X_c\,\ell\,
\bar\nu$ decays, the lepton spectrum can be written as a convolution
of the free-quark decay distribution with the universal structure
function $f(k_+)$. We have shown that the effect of the momentum
distribution of the heavy quark inside the meson can be understood in
terms of an effective mass $m_b^*=m_b+k_+$, which determines the
decay kinematics. Over most of phase space, the free-quark decay
distribution is slowly varying on scales of order $\Lambda$, whereas
the light-cone distribution function is sharply peaked around
$k_+\sim 0$ with an intrinsic width of order $\Lambda$. In this case,
the convolution reproduces the parton model up to small
nonperturbative corrections of order $(\Lambda/m_b)^2$. In the
endpoint region, however, the free-quark decay distribution falls
steeply, and the convolution becomes sensitive to the details of the
structure function. This leads to large, genuinely nonperturbative
effects in the lepton spectrum, which may all be attributed to
bound-state corrections in the initial state. In particular, as one
approaches the endpoint, the effective mass approaches the mass of
the physical $B$-meson, and one recovers the correct position of the
maximum lepton energy.

We have illustrated the effects of nonperturbative corrections using
a simple one-parameter model, which however includes many of the
ingrediences of a more sophisticated description. For simplicity, and
since our main focus in this paper was to investigate the effects of
bound-state corrections, we have not included in our discussion
perturbative QCD corrections from the emission of real and virtual
gluons. Such effects have been considered in
Refs.~\cite{AlPi,CCM,JeKu}, and more recently in
Refs.~\cite{Fermi,AJMW}. For inclusive decays into final states
containing light quarks, their interplay with the nonperturbative
corrections considered here is rather intricate because of the
presence of large Sudakov double-logarithms. In particular, in $B\to
X_u\,\ell\,\bar\nu$ decays the simple factorization of bound-state
corrections into an integral over $f(k_+)$ [see (\ref{buconv})] is
replaced by a more complicated convolution of the structure function
with a hard-scattering amplitude. For the case of $B\to
X_c\,\ell\,\bar\nu$ decays, however, such effects are known to be
less severe. Nevertheless, further investigation of radiative
corrections is necessary to put our formalism on a more quantitative
basis.

In the last part of the paper, we have discussed some possible
approaches that could be taken to extract information about the
structure function from semileptonic decay spectra. We are aware that
such an analysis will be complicated, due to various theoretical and
experimental limitations. It is likely that the most promising
approach will be to make a theory-motivated ansatz for the structure
function that contains few physical parameters, and to extract these
parameters from a fit to data. We are confident that it should be
possible to obtain an (approximate) determination of the first two
non-trivial moments of $f(k_+)$. They are proportional to the average
kinetic energy of the $b$-quark inside the $B$-meson, and to the
asymmetry parameter $A_3$. These fundamental parameters are of
sufficient interest to try such an analysis using the existing data
on $B\to X_c\,\ell\,\bar\nu$ decays.

\bigskip\bigskip
\noindent
{\it Acknowledgements:\/}
It is a pleasure to thank Daniel Wyler and Matthias Jamin for useful
discussions.

\newpage

\end{document}